%% file: orbits.tex
\documentclass[twocolumn]{aastex631}
\usepackage{graphicx}
\usepackage{placeins}


\usepackage{xcolor}
\usepackage{soul}  

\begin{document}

\title{New Orbital Constraints for YSES 1 b and HR 2562 B from High-Precision Astrometry and Planetary Radial Velocities\footnote{Based on observations collected at the European Southern Observatory under ESO programmes 109.238N.002, 110.23U4.001, 109.238N.004, 113.26QU.001, 1104.C-0651(B), 109.238N.002, 109.238N.004, 60.A-9102(J), and 114.27UV.001}}

\author[0000-0003-1278-4722]{Jonathan Roberts}
\affiliation{Center for Interdisciplinary Exploration and Research in Astrophysics (CIERA), Northwestern University, 1800 Sherman Avenue, Evanston, IL 60201, USA}
\affiliation{Department of Physics and Astronomy, Northwestern University, 2145 Sheridan Road, Evanston, IL 60208, USA}

\author[0000-0001-5684-4593]{William Thompson}
\affiliation{NRC Herzberg Astronomy and Astrophysics, 5071 West Saanich Road,Victoria, BC, V9E 2E7, Canada}

\author[0000-0003-0774-6502]{Jason~J.~Wang}
\affiliation{Center for Interdisciplinary Exploration and Research in Astrophysics (CIERA), Northwestern University, 1800 Sherman Avenue, Evanston, IL 60201, USA}
\affiliation{Department of Physics and Astronomy, Northwestern University, 2145 Sheridan Road, Evanston, IL 60208, USA}

\author[0000-0002-3199-2888]{Sarah Blunt}
\affiliation{Department of Astronomy \& Astrophysics, University of California, Santa Cruz, Santa Cruz, CA, USA, 95060}
\affiliation{Center for Interdisciplinary Exploration and Research in Astrophysics (CIERA), Northwestern University, Evanston, IL 60208, USA}
\affiliation{NSF Astronomy and Astrophysics Postdoctoral Fellow}

\author[0000-0001-6396-8439]{William O. Balmer}
\affiliation{Department of Physics \& Astronomy, Johns Hopkins University, 3400 N. Charles Street, Baltimore, MD 21218, USA}
\affiliation{Space Telescope Science Institute, 3700 San Martin Drive, Baltimore, MD 21218, USA}

\author{Guillaume Bourdarot}
\affiliation{Max Planck Institute for Extraterrestrial Physics, Giessenbachstrasse, 85748 Garching, Germany}

\author[0000-0003-2649-2288]{Brendan P. Bowler}
\affiliation{Department of Physics, University of California, Santa Barbara, Santa Barbara, CA 93106, USA}

\author{Gael Chauvin}
\affiliation{Laboratoire Lagrange, UniversiteĆote d'Azur, CNRS, Observatoire de la Cote d'Azur, 06304 Nice, France}

\author{Frank Eisenhauer}
\affiliation{Max Planck Institute for Extraterrestrial Physics, Giessenbachstrasse, 85748 Garching, Germany}

\author{Thomas K. Henning}
\affiliation{Max-Planck-Institut für Astronomie, Königstuhl 17, 69117 Heidelberg, Germany}

\author[0000-0003-2769-0438]{Jens Kammerer}
\affiliation{European Southern Observatory, Karl-Schwarzschild-Straße 2, 85748 Garching, Germany}

\author{Flavien Kiefer}
\affiliation{LIRA, CNRS, Observatoire de Paris, CNRS, Université PSL, 5 Place Jules Janssen, 92190 Meudon, France}

\author[0000-0002-7064-8270]{Matthew A. Kenworthy}
\affiliation{Leiden Observatory, Leiden University, Postbus 9513, 2300 RA Leiden, The Netherlands}

\author[0000-0003-0626-1749]{Pierre Kervella}
\affiliation{LIRA, Observatoire de Paris, Université PSL, Sorbonne Université, Université Paris Cité, CNRS, 92190 Meudon, France}
\affiliation{French-Chilean Laboratory for Astronomy, IRL 3386, CNRS and U. de Chile, Casilla 36-D, Santiago, Chile}

\author[0000-0002-6948-0263]{Sylvestre Lacour}
\affiliation{LIRA, Observatoire de Paris, Université PSL, Sorbonne Université, Université Paris Cité, CNRS, 92190 Meudon, France}

\author{A.-M. Lagrange} 
\affiliation{LIRA, CNRS, Observatoire de Paris, CNRS, Université PSL, 5 Place Jules Janssen, 92190 Meudon, France}  
\affiliation{Univ. Grenoble Alpes, CNRS, IPAG, F-38000 Grenoble, France}

\author[0000-0001-6975-9056]{Eric L. Nielsen}
\affiliation{New Mexico State University, 1320 Frenger Mall, Las Cruces, NM 88003 USA }

\author[0000-0003-3818-408X]{Laurent Pueyo}
\affiliation{Space Telescope Science Institute, 3700 San Martin Drive, Baltimore, MD 21218, USA}

\author[0000-0003-4203-9715]{Emily Rickman}
\affiliation{European Space Agency (ESA), ESA Office, Space Telescope Science Institute, 3700 San Martin Drive, Baltimore, MD 21218, USA}

\author{Olli Sipilä}
\affiliation{Max Planck Institute for Extraterrestrial Physics, Giessenbachstrasse, 85748 Garching, Germany}

\author{Silvia Spezzano}
\affiliation{Max Planck Institute for Extraterrestrial Physics, Giessenbachstrasse, 85748 Garching, Germany}

\author[0000-0002-5823-3072]{Tomas Stolker}
\affiliation{Leiden Observatory, University of Leiden, PO Box 9513, 2300 RA Leiden, The Netherlands}

\author[0000-0002-5903-8316]{Alice Zurlo}
\affiliation{Instituto de Estudios Astrof\'isicos, Facultad de Ingenier\'ia y Ciencias, Universidad Diego Portales, Av. Ej\'ercito Libertador 441, Santiago, Chile}
\affiliation{Millennium Nucleus on Young Exoplanets and their Moons (YEMS)}

\begin{abstract}

    We present new VLTI/GRAVITY astrometry and updated orbit fits for the directly imaged companions YSES 1 b and HR 2562 B, substellar objects straddling the planet-brown dwarf boundary. Using high-precision astrometry, radial velocity (RV) data, and proper motions, we derive revised orbital parameters with \texttt{orbitize!} \citep{Blunt_2020}. For YSES 1 b, the inclusion of GRAVITY astrometry and a relative radial velocity measurement from \citet{zhang2024} overcomes the traditional challenge of constraining eccentricities for distant companions, enabling the first orbit fit and yielding a constrained eccentricity of $0.44\pm0.20$. This represents the first full orbit fit for the system. Additionally, we calculate a median line-of-sight stellar obliquity of $12^{+11}_{-8}$ deg, providing further insight into the system's dynamical architecture. For HR 2562 B, our analysis agrees with \citet{Zhang_2023}, confirming a low-eccentricity orbit ($0.34\pm0.20$) and an inclination of $87\pm 1$ deg. We find HR 2562 B's orbit to be nearly coplanar with the debris disk, with a mutual inclination of $3.7\pm 0.3$ deg. For both YSES 1 b and HR 2562 B the lower eccentricities favor an {in situ} formation scenario over extreme scattering or cloud fragmentation.

\end{abstract}


\section{Introduction}

Direct imaging of exoplanets and substellar objects, such as brown dwarfs and super-Jupiters, has significantly advanced our understanding of the formation and evolution of planetary systems. For instance, the direct imaging of HR 8799's multi-planet system revealed four giant exoplanets orbiting their host star at large separations, challenging traditional models of planet formation via core accretion \citep{ Marois_2010, Wang_2018}. Observations of $\beta$ Pictoris b and c, a directly imaged planet pair in a young debris disk, have provided a unique opportunity to study planet-disk interactions and the dynamical evolution of young systems \citep{Lagrange_2010,AML_2019,nowak_19}. Similarly, the PDS 70 system hosts two protoplanets within a transition disk, including the first detection of accreting planets and circumplanetary material \citep{Keppler_2018, haffert_2019}, highlighting the power of multi-planet imaging in tracing the early stages of system formation.

Two specific objects, YSES 1 b and HR 2562 B, provide insight into the boundary between giant planets and brown dwarfs, a distinction often defined by the deuterium-burning limit of approximately 13 $M_\mathrm{Jup}$. YSES 1 b orbits a $17 \pm 1$ Myr K3IV star at a distance of $95 \pm 1$ pc \citep{bailer_jones_2018, Gaia_collab_2018}, while HR 2562 B orbits an F5V star with an age of 300–900 Myr at $34 \pm 1$ pc \citep{KONO2016}. YSES 1 b with a model-independent mass estimate of approximately $14 \pm3$ $M_\mathrm{Jup}$ \citep{Bohn_2019, Bohn_2020}, straddles this threshold and exhibits atmospheric characteristics of a low-gravity, young object. YSES 1 c, an object with an estimated mass of approximately $6$ $M_\mathrm{Jup}$ \citep{Bohn_2020}, further enriches the system's complexity, as its 320 AU \citep{Bohn_2020} separation from YSES 1 b raises questions about its formation and dynamical interactions. An orbital fit for YSES 1 c was not performed, as we lack sufficient astrometric measurements to meaningfully constrain its orbital parameters. Continued astrometric monitoring may eventually enable a dynamical analysis of both companions. In this work, we present updated astrometry and orbit fits for YSES 1 b as a step toward a more complete dynamical picture of the system. Determining their orbital eccentricities will be key to distinguishing between these scenarios; lower eccentricities might favor \textit{in situ} formation, while higher eccentricities could indicate extreme scattering. As \citet{Bowler_2020} and \citet{Vighnesh_23} highlight (and references therein), the orbital eccentricities of these objects can serve as a fossil record of their dynamical past. While no clear resonances have been identified, the presence of multiple companions in a young system like this invites comparisons to systems such as HR 8799, where multi-companion dynamics—including resonance chains—play a critical role in the system's architecture and stability \citep{Bohn_2020,Alice_2022}. 

In the case of HR 2562, its debris disk, first observed with Herschel \citep{KONO2016}, and subsequent monitoring with instruments like VLT/SPHERE, has revealed a system architecture consistent with a coplanar geometry between the companion and the disk \citep{Maire_2018}. The system's poorly constrained age of $300-900  $Myr leads to a broad mass range of $15-45 $ $M_\mathrm{Jup}$, depending on the evolutionary model \citep{KONO2016, Zhang_2023}. This range places HR 2562 B near the boundary between giant planets and brown dwarfs. Residing within 30 AU of its host star, HR 2562 B offers a complementary case for studying substellar objects within debris disks and refining our understanding of this transition zone.

In this work, we present new GRAVITY observations for YSES 1 b and HR 2562 B. In Section 2, we describe the observations and data reduction process, including details about the GRAVITY instrument and the observational setup. Through these analyses, we aim to address several key questions: Can the GRAVITY data provide more precise constraints on the inclination and eccentricity of their orbits, helping to refine our understanding of their formation? In Section 3, we detail the orbit analysis, performed using \texttt{orbitize!}, focusing on eccentricity, inclination, and potential coplanarity with circumstellar disks. In Section 4, we present the results of these analyses. Finally, in Section 5, we discuss the implications of our findings, emphasizing the importance of additional data and complementary simulations to further refine our understanding of their formation and dynamics.

\section{Observations and Data}

\subsection{GRAVITY Observations}

In this work, we obtain and analyze new astrometry on YSES 1 b and HR 2562 B using the GRAVITY \citep{GravityCollaboration:2017a} instrument on the Very Large Telescope Interferometer (VLTI) at the European Southern Observatory (ESO), enabling us to refine their orbital parameters and better understand their dynamical properties. The VLTI operates with either four 8.2-meter Unit Telescopes (UTs) or four 1.8-meter Auxiliary Telescopes (ATs), enabling high-resolution infrared interferometry. In this work, we present observations that utilized both the ATs and UTs. GRAVITY operates in the K-band (2.0–2.4 µm) and uses dual-field interferometry, a technique that allows a bright reference star to be used for fringe tracking, enabling simultaneous observations of a faint science target. For these observations, GRAVITY was operated in dual-field mode, which allows for precise phase-referenced interferometry by observing the science target along with a nearby reference star to correct atmospheric and instrumental phase errors. In our cases, the reference star was the host star of the companion. GRAVITY was used in both on-axis and off-axis configurations during our observations. The targets observed in this study are the directly imaged objects YSES 1 b and HR 2562 B. Observations of YSES 1 b were conducted eight times, on 2023-01-27, 2024-02-15, 2024-05-30,  2024-05-31, 2024-06-23, 2024-06-24, 2024-06-25, and 2024-06-30. HR 2562 B was observed five times, on 2022-01-25, 2022-10-24, 2023-10-04, 2023-10-12, and 2024-03-19. All dates are in UT. These observations were part of the ExoGRAVITY program ID's 109.238N.002, 110.23U4.001, 109.238N.004, 113.26QU.001, 1104.C-0651(B), 109.238N.002, 109.238N.004, 60.A-9102(J), and 114.27UV.001. 

To calibrate astrometry for off-axis mode \citep{Lacour_2020}, binary stars found from \citet{nowak_ref} were used as phase reference metrology calibrators. For HR 2562 B, the reference stars HD 73900 and HD 30003A were employed, while for YSES 1 b, the reference star used were HR 5362 and HD 123227 A. A comprehensive observing log detailing each observation date, exposure time, and observing conditions is provided in Table \ref{tab:observing_log}.

\begin{deluxetable*}{ccccccccc}
\tablewidth{\textwidth}
\tablecaption{Observing log for HR 2562 B and YSES 1 b\label{tab:observing_log}}
\tablehead{Target & Date & Start & End & NEXP/NDIT/DIT & Airmass & $\tau_0$ & Seeing & Fiber pointing \\
& UT date & UT time & UT time &  & & & & $\Delta$RA/$\Delta$DEC}
\startdata
HR 2562 & 2022-01-25 & 05:49:53 & 05:57:10 & 2/4/100s & 1.40--1.44 & 5.8--7.3 ms & 0.62--0.75$^{\prime\prime}$ & [-651.1, 362.0] \\
HD 73900 & 2022-01-25 & 06:15:19 & 06:17:41 & 3/96/0.3s & 1.06--1.06 & 5.9--6.7 ms & 0.52--0.68$^{\prime\prime}$ & [-826.635, -456.09] \\
HD 73900 & 2022-01-25 & 06:21:59 & 06:23:12 & 2/96/0.3s & 1.07--1.07 & 6.6--7.6 ms & 0.57--0.62$^{\prime\prime}$ & [826.635, 456.09] \\
HR 2562 B & 2022-10-24 & 06:39:16 & 08:30:04 & 7/4/100s & 1.25--1.42 & 2.0--6.1 ms & 0.47--0.92$^{\prime\prime}$ & [-669.4, 370.8] \\
HR 2562 & 2022-10-24 & 06:34:12 & 08:27:17 & 9/8/10s & 1.25--1.43 & 2.0--6.1 ms & 0.46--0.95$^{\prime\prime}$ & [0.0, 0.0] \\
HR 2562 B & 2023-10-04 & 08:29:34 & 09:06:50 & 3/4/100s & 1.29--1.35 & 2.9--3.9 ms & 0.56--0.63$^{\prime\prime}$ & [-685.4, 381.9] \\
HD 30003 A & 2023-10-04 & 07:48:04 & 07:49:00 & 2/8/3s & 1.22--1.22 & 3.7--34.2 ms & 0.47--0.50$^{\prime\prime}$ & [-3953.4, -41.1] \\
HR 2562 B & 2023-10-12 & 08:28:51 & 09:06:01 & 3/4/100s & 1.25--1.30 & 5.1--6.3 ms & 0.44--0.52$^{\prime\prime}$ & [-685.3, 382.1] \\
HR 2562 & 2023-10-12 & 08:24:03 & 09:08:25 & 4/8/10s & 1.25--1.30 & 4.7--6.8 ms & 0.44--0.55$^{\prime\prime}$ & [0.0, 0.0] \\
HR 2562 B & 2024-03-19 & 04:17:22 & 04:55:51 & 4/4/100s & 1.89--2.19 & 3.3--5.3 ms & 0.44--0.76$^{\prime\prime}$ & [-688.07, 385.46] \\
HR 2562 & 2024-03-19 & 04:13:09 & 04:58:00 & 3/8/10s & 1.86--2.21 & 3.3--4.9 ms & 0.53--0.76$^{\prime\prime}$ & [0.0, 0.0] \\
\hline
YSES 1 b & 2023-01-27 & 07:33:58 & 08:29:20 & 5/4/100s & 1.35--1.42 & 5.4--7.9 ms & 0.70--1.14$^{\prime\prime}$ & [-905.6, -1449.3] \\
HR 5362 & 2023-01-27 & 07:13:49 & 07:15:50 & 2/4/10s & 1.42--1.42 & 8.1--8.7 ms & 0.70--0.89$^{\prime\prime}$ & [-2944.2  1750.1] \\
HR 5362 & 2023-01-27 & 07:21:37 & 07:22:53 & 2/16/1s & 1.38--1.39 & 7.3--8.0 ms & 0.87--1.89$^{\prime\prime}$ & [2944.2  -1750.1] \\
YSES 1 b & 2024-02-15 & 06:12:13 & 07:17:07 & 4/4/100s & 1.35--1.44 & 10.8--18.9 ms & 0.39--0.64$^{\prime\prime}$ & [-905.6, -1449.3] \\
HR 5362 & 2024-02-15 & 05:51:08 & 06:42:28 & 4/4/10s & 1.27--1.47 & 10.1--18.9 ms & 0.39--0.51$^{\prime\prime}$ & [-2941.9, 1747.3] \\
HR 5362 & 2024-02-15 & 05:57:44 & 06:49:43 & 4/16/1s & 1.25--1.47 & 10.4--18.9 ms & 0.41--0.51$^{\prime\prime}$ & [2941.9, -1747.3] \\
YSES 1 b & 2024-05-30 & 00:05:40 & 01:06:36 & 6/4/100s & 1.32--1.37 & 3.6--4.7 ms & 0.62--0.96$^{\prime\prime}$ & [-906.0, -1434.0] \\
HD 123227 A & 2024-05-29 & 23:43:29 & 23:52:43 & 4/120/0.30s & 1.27--1.29 & 4.3--5.4 ms & 0.58--0.71$^{\prime\prime}$ & [378.11, -882.83] \\
YSES 1 b & 2024-05-31 & 00:16:56 & 00:24:11 & 2/4/100s & 1.34--1.35 & 3.7--4.2 ms & 0.58--0.70$^{\prime\prime}$ & [-906.0, -1434.0] \\
HD 123227 A & 2024-05-30 & 23:24:35 & 23:32:53 & 4/120/0.30s & 1.31--1.33 & 2.8--3.8 ms & 0.76--1.12$^{\prime\prime}$ & [378.11, -882.83] \\
YSES 1 b & 2024-06-23 & 00:21:37 & 01:18:00 & 6/4/100s & 1.31--1.36 & 2.3--2.3 ms & 2.10--2.10$^{\prime\prime}$ & [-906.0, -1434.0] \\
HD 123227 A & 2024-06-22 & 23:23:31 & 00:05:18 & 6/120/0.30s & 1.12--1.15 & 2.3--2.3 ms & 2.10--2.10$^{\prime\prime}$ & [378.11, -882.83] \\
YSES 1 b & 2024-06-24 & 23:33:31 & 00:19:33 & 4/4/100s & 1.31--1.32 & 2.0--2.5 ms & 0.69--0.85$^{\prime\prime}$ & [-906.0, -1434.0] \\
HD 123227 A & 2024-06-24 & 23:11:47 & 23:20:20 & 4/120/0.30s & 1.15--1.60 & 2.3--2.3 ms & 1.97--1.97$^{\prime\prime}$ & [378.11, -882.83] \\
YSES 1 b & 2024-06-25 & 23:59:05 & 00:50:16 & 6/4/100s & 1.31--1.34 & 2.0--3.0 ms & 0.59--0.82$^{\prime\prime}$ & [-906.0, -1434.0] \\
HD 123227 A & 2024-06-25 & 23:34:19 & 23:47:01 & 4/120/0.30s & 1.12--1.13 & 1.6--3.5 ms & 0.49--0.74$^{\prime\prime}$ & [378.11, -882.83] \\
YSES 1 b & 2024-06-30 & 00:16:20 & 02:11:00 & 12/4/100s & 1.32--1.46 & 4.0--7.0 ms & 0.56--0.80$^{\prime\prime}$ & [-906.0, -1434.0] \\
HD 123227 A & 2024-06-29 & 23:56:00 & 02:28:00 & 4/120/0.30s & 1.11--1.25 & 3.3--6.9 ms & 0.56--0.85$^{\prime\prime}$ & [-378.11, 882.83] \\
HD 123227 A & 2024-06-30 & 00:03:09 & 02:34:27 & 4/120/0.30s & 1.11--1.25 & 4.0--6.3 ms & 0.58--0.69$^{\prime\prime}$ & [378.11, -882.83] \\
\enddata
\end{deluxetable*}

\begin{deluxetable*}{ccccccc}
    \tablewidth{\textwidth}
    \tablecaption{New relative astrometry presented in this paper. $\sigma_{\rm R.A.}$ and $\sigma_{ \rm Dec.}$ denote the uncertainties in astrometric position, and $\rho_{\rm R.A.,  Decl.}$ denotes the correlation between the $\sigma_{\rm R.A.}$ and $\sigma_{\rm Dec.}$ measurements.
    \label{tab:astrom}}
    \tablehead{  Object & Date & R.A. & $\sigma_{\rm R.A.}$ & Decl. & $\sigma_{\rm Decl.}$ & $\rho_{\rm R.A., Decl.}$ \\ & $[\mathrm{JD} - 2400000.5]$ & [mas] & [mas] & [mas] & [mas]& [mas] }
    \startdata
    HR 2562 B & 59604.250 & -653.980 & 0.030 & 360.350 & 0.067 & 0.368 \\
    HR 2562 B & 59876.320 & -663.190 & 0.176 & 366.660 & 0.244 & -0.460 \\
    HR 2562 B & 60221.370 & -672.190 & 0.237 & 374.480 & 0.100 & -0.968 \\
    HR 2562 B & 60229.370 & -672.780 & 0.072 & 375.010 & 0.109 & -0.070 \\
    HR 2562 B & 60388.000 & -676.790 & 0.829 & 377.760 & 0.301 & 0.050 \\
    \hline
    YSES 1 b & 59971.334 & -906.598 & 0.053 & -1436.899 & 0.055 & -0.629 \\
    YSES 1 b & 60355.278 & -904.806 & 0.068 & -1434.773 & 0.137 & -0.576 \\
    YSES 1 b & 60460.026 & -904.515 & 0.057 & -1434.091 & 0.048 & -0.804 \\
    YSES 1 b & 60461.014 & -904.428 & 0.094 & -1434.237 & 0.103 & -0.901 \\
    YSES 1 b & 60484.032 & -904.440 & 0.021 & -1433.937 & 0.026 & -0.121 \\
    YSES 1 b & 60485.996 & -904.480 & 0.045 & -1433.821 & 0.093 & -0.876 \\
    YSES 1 b & 60487.016 & -904.455 & 0.030 & -1433.970 & 0.043 & -0.590 \\
    YSES 1 b & 60491.050 & -904.516 & 0.070 & -1433.772 & 0.120 & 0.904  \\   \enddata
\end{deluxetable*}

\subsection{GRAVITY Data Reduction}
For both objects, we calibrated the raw data using the ESO GRAVITY pipeline \citep{Lapeyrere:2014a}, Public Release 1.6.6 (1 March 2024\footnote{\url{https://www.eso.org/sci/software/pipelines/gravity}}).

We employed the open source \texttt{exogravity} pipeline\footnote{\url{https://gitlab.obspm.fr/mnowak/exogravity}} to isolate the planet's flux from the host star's light. The calibrator binaries were used to measure the metrology zero point and phase reference the observations of the planets \citep{nowak_ref}. Then, as described in Appendix A of \citet{GravityCollaboration:2020a}, the planet's signal and stellar contamination are modeled as a combination of a point source signal and a time- and baseline-dependent polynomial function.

To determine the astrometric position of the planet relative to its host star at each epoch, we computed a $\chi ^2$ map across the fiber's field-of-view. This map allowed us to identify the location of the planet's signal. Following the approach outlined in \citet{Blunt_2023}, we determined the planet's mean astrometric position by identifying the point in the map with the minimum chi-square ($\chi^2$) value, which represents the best-fit location of the planet relative to the star.

We assessed the uncertainty of each astrometric measurement by examining the scatter in mean astrometric values across individual exposures, following the methodologies of \citet{gravitycollab_2019,GravityCollaboration:2020a}. This approach provides a robust estimate of measurement error, as it accounts for random noise and systematic variations within individual exposures. The resulting new astrometric positions for the planet are presented in Table \ref{tab:astrom}. We assumed a constant contrast when extracting the astrometry.

\subsection{Literature Data}

In this work, we supplemented our GRAVITY observations with all available archival data from the Gemini Planet Imager \citep[GPI;][]{KONO2016,Zhang_2023}, SPHERE/IRDIS \citep{Zhang_2023}, and NACO \citep{BOHN2019}, providing complementary astrometric information. All literature data used is located in Table \ref{tab:lit_astrom}. 

For both HR 2562 B and YSES 1 b, the GRAVITY astrometric measurements significantly improved the precision of previous literature values. The Right Ascension (R.A.) and the Declination (Decl.) uncertainties were reduced by 10$\times$ compared to earlier data. These high-precision measurements were critical inputs for the orbit fitting process. For YSES 1b, we incorporated a relative radial velocity (RV) measurement from \citet{zhang2024}, which provides an additional constraint on the orbital motion along the line of sight. This relative RV measurement represents the velocity offset between YSES 1b and its host star, measured to be $-1.87 \pm 0.04$ km s$^{-1}$ using CRIRES+ observations taken on 2023 February 27 and 28. This relative RV helps break degeneracies in orbital inclination and true mass, and is included directly in our orbital fit. These precise GRAVITY and RV measurements form the foundation of our refined orbital analysis, which is presented in the next section. In our orbit analysis for HR 2562B, we incorporated Hipparcos-Gaia Catalog of Accelerations \citep[HGCA;][]{brandt_2018, brandt_2021} proper motions, which provided additional constraints on the orbital fit by helping to refine the system's dynamical motion.

\begin{deluxetable*}{cccccccc}
    \tablewidth{\textwidth}
    \tablecaption{Astrometric data from the literature used in this work
    \label{tab:lit_astrom}}
    \tablehead{Object& Date & R.A. & $\sigma_{R.A.}$ & Decl. & $\sigma_{Decl.}$ & Instrument & Reference\\ &$[\mathrm{JD} - 2400000.5]$ & [mas] & [mas] & [mas] & [mas] &  & }
\startdata
HR 2562 B &57412.1335	&	-543.9113	&	2.0626	&	287.1359	&	2.6915	&	Gemini-S/GPI & (1) \\
HR 2562 B &57415.1731	&	-542.0582	&	1.7937	&	285.3114	&	2.4609	&	Gemini-S/GPI & (1) \\
HR 2562 B &57415.2002	&	-541.2088	&	1.6547	&	288.8572	&	2.3459	&	Gemini-S/GPI & (1) \\
HR 2562 B &57443.0343	&	-545.4322	&	1.6206	&	286.4797	&	2.3557	&	Gemini-S/GPI & (1) \\
HR 2562 B &57446.0951	&	-545.3914	&	1.7438	&	287.3090	&	2.4633	&	Gemini-S/GPI & (1) \\
HR 2562 B &57734.2647	&	-564.1338	&	6.3161	&	297.5599	&	6.0988	&	SPHERE/IRDIS & (2) \\
HR 2562 B &57739.3123	&	-565.0942	&	1.4324	&	300.2129	&	1.7791	&	Gemini-S/GPI & (1) \\
HR 2562 B &57791.1114	&	-569.5653	&	2.2646	&	300.5518	&	2.1719	&	SPHERE/IRDIS & (2) \\
HR 2562 B &57797.0517	&	-567.5762	&	1.3716	&	304.7144	&	1.6868	&	Gemini-S/GPI & (1) \\
HR 2562 B &58025.3826	&	-583.9674	&	1.4380	&	310.1089	&	1.7409	&	SPHERE/IRDIS & (2) \\
HR 2562 B &58025.3826	&	-581.3417	&	1.6852	&	310.1468	&	1.8825	&	SPHERE/IRDIS & (2) \\
HR 2562 B &58086.3110	&	-584.9734	&	2.2305	&	315.8982	&	2.6348	&	Gemini-S/GPI & (1) \\
HR 2562 B &58149.1962	&	-588.0363	&	1.5599	&	319.9424	&	2.1367	&	Gemini-S/GPI & (1) \\
HR 2562 B &58187.0475	&	-588.1997	&	2.8811	&	322.5639	&	2.9963	&	Gemini-S/GPI & (1) \\
HR 2562 B &58441.3261	&	-600.4164	&	1.3278	&	331.3109	&	1.4902	&	Gemini-S/GPI & (1) \\
\hline
YSES 1 b &57939.00	&	-911.3486	&	5.4996	&	-1452.8141	&	4.6487	&	SPHERE/IRDIS	& (3) \\
YSES 1 b &58559.00	&	-906.5660	&	7.8620	&	-1445.1900	&	5.8337	&	SPHERE/IRDIS	& (3) \\
YSES 1 b &58565.00	&	-906.6919	&	2.9896	&	-1451.0103	&	2.9959	&	SPHERE/IRDIS	& (3) \\
YSES 1 b &58621.00	&	-920.2205	&	5.6976	&	-1438.9087	&	5.2978	&	NACO	& (3) \\
YSES 1 b &58637.00	&	-917.8713	&	11.9728	&	-1446.3337	&	11.9909	&	NACO	& (3) \\
YSES 1 b &58895.00	&	-907.6288	&	2.9860	&	-1446.8842	&	2.9952	&	SPHERE/IRDIS	& (4) \\
\enddata
\tablecomments{References: (1)~\citet{Zhang_2023},
(2)~\citet{Maire_2018},
(3)~\citet{BOHN2019}.,
(4)~\citet{Bohn_2020}.
}
\end{deluxetable*}

\section{ORBIT ANALYSIS}

Orbit fits are performed using the \texttt{orbitize!} package \citep{Blunt_2020}, which supports a range of sampling algorithms for constraining the orbital parameters of directly imaged companions. For this analysis, we employed the \texttt{ptemcee} parallel-tempered Markov Chain Monte Carlo (MCMC) sampler \citep{FM_2013, V_2016}, which is particularly well suited for exploring complex, multimodal posterior distributions. Our fits used 20 temperatures and 1000 walkers per temperature, with an initial burn-in phase of 100,000 steps to ensure convergence. 

The orbital fits incorporate astrometric measurements and radial velocities, along with Hipparcos-Gaia Catalog of Accelerations proper motions for HR 2562. Stellar mass and parallax are treated as free parameters, with priors centered on the respective catalog values. A full description of the prior distributions adopted for the orbital parameters is provided in Table \ref{tab:posterior_results}.
\begin{figure*}[t]
    \centering
    \includegraphics[width=\linewidth]{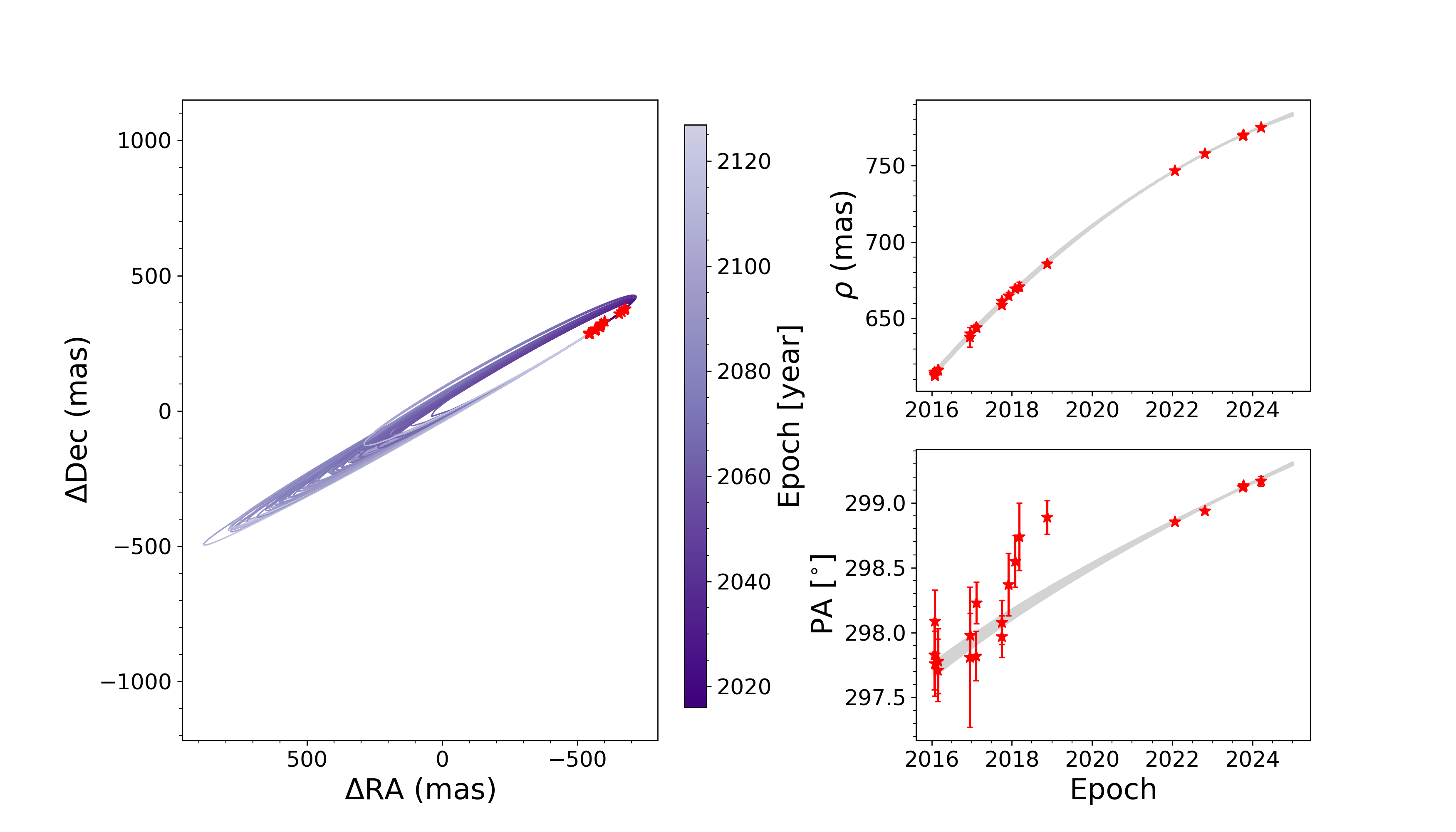}
    \caption{
Orbit fit for HR 2562 B. The leftmost plot displays 100 randomly selected orbits, color-coded to illustrate the anticipated orbital position over time. The right panels present the measured separation and position angle as a function of time, compared to the randomly drawn orbits (in gray). Red stars represent the observations. The apparent position angle discrepancy observed between 2018 to 2020 is likely due to systematic effects inherent to GPI astrometry. }
    \label{fig:hr2562_vis}
\end{figure*}
\begin{figure*}[t]
    \centering
    \includegraphics[width=0.7\linewidth]{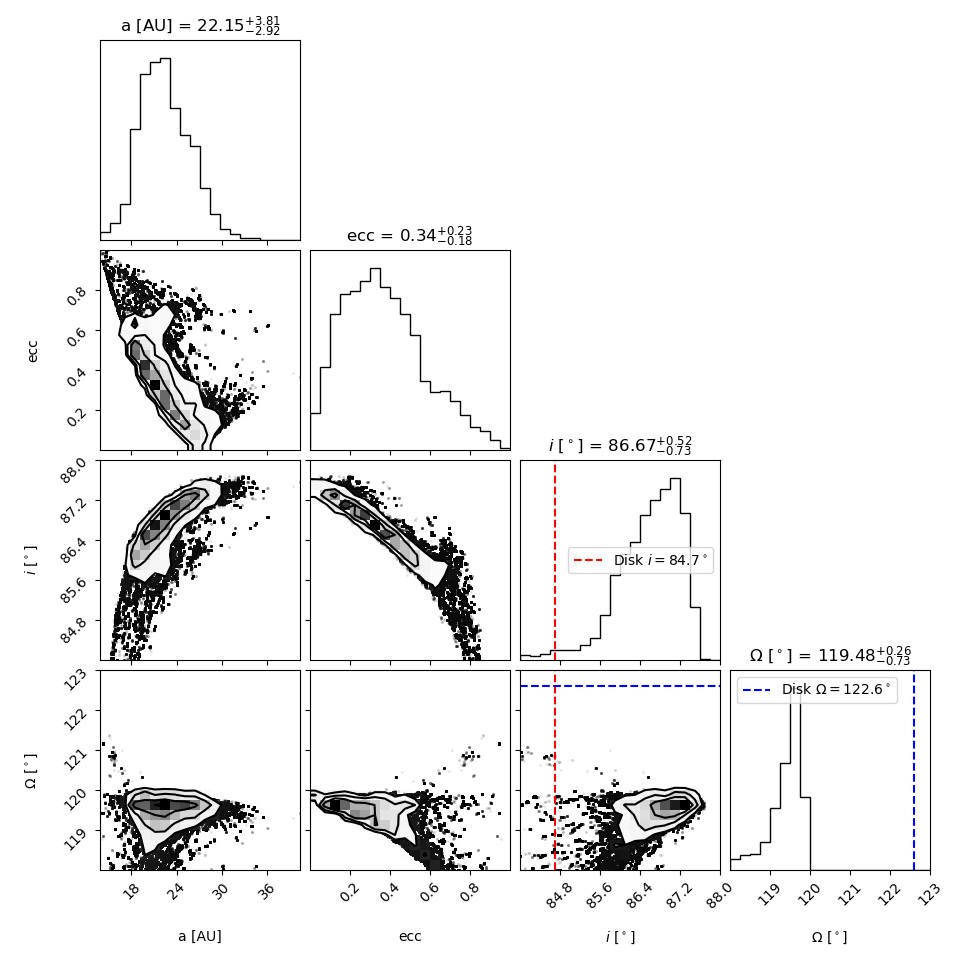}
    \caption{Corner plot for the orbital fit of HR 2562 B. The object demonstrates a preference for eccentric orbits less than 0.4, with the eccentricity parameter favoring values closer to circular. The semi-major axis distribution shows a prominent peak around 22 AU, with no values less than 14 AU. The orientation of the disk around HR 2562 B is nearly edge-on, as reflected in the inclination distribution, which supports this coplanar alignment between the orbit and the disk.}
    \label{fig:HR2562_small}
\end{figure*}

For YSES 1 b, we performed an orbit fit using a subset of four astrometric epochs, approximately one per month. These epochs were chosen to be MJD 59971.332, 60355.278, 60461.014, and 60491.050. This selection was made to minimize the influence of correlated measurements taken within days of each other, which can lead to significant variations in the accepted orbits. By selecting the epochs with the highest error, we preserve the orbit while reducing the impact of short-timescale systematics.

\begin{figure*}[t]
    \centering
    \includegraphics[width=\linewidth]{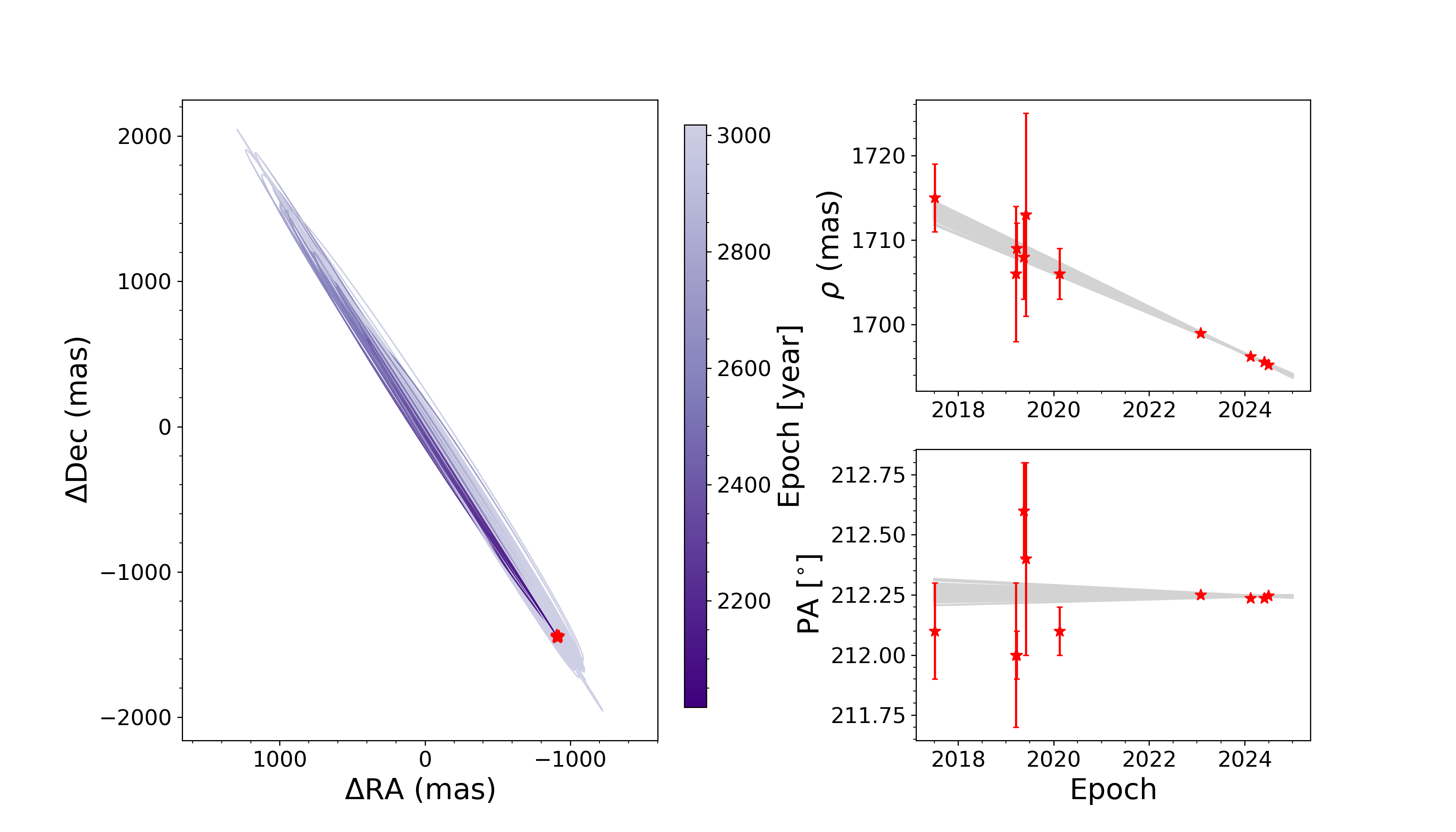}
    \caption{
    Orbit fit for YSES 1 b. See \ref{fig:hr2562_vis} for details.}
    \label{fig:yses_vis}
\end{figure*}

The final posterior samples are used to compute marginalized constraints on the orbital elements and generate visualizations of the fitted orbits, as shown in Table \ref{tab:posterior_results} summarizes the priors used in our analysis. The high-precision astrometry plays a key role in tightening constraints on the orbital geometry, particularly for parameters such as inclination and longitude of the ascending node.

\begin{figure}[h]
    \centering
    \includegraphics[width=\linewidth]{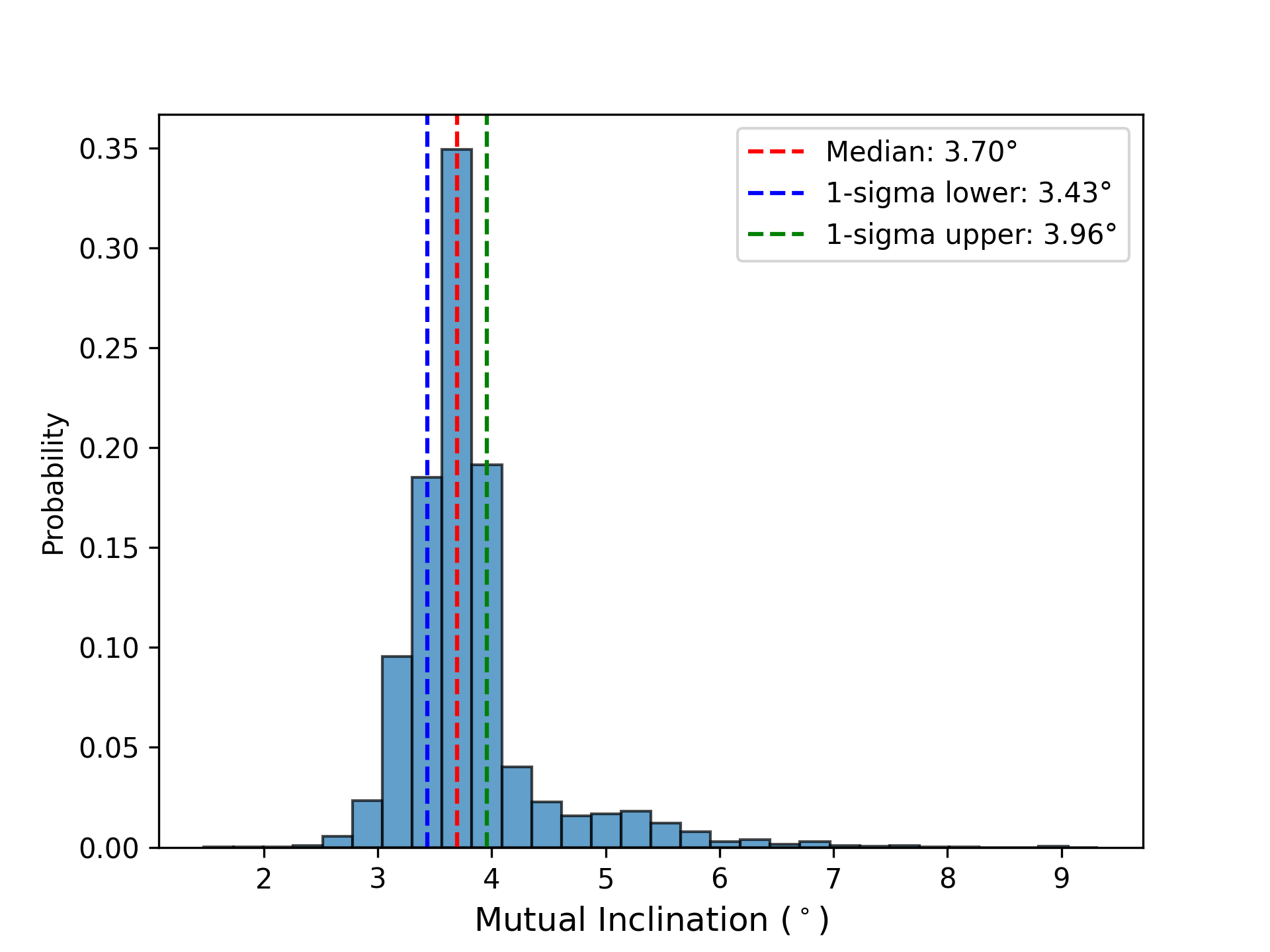}
    \caption{Histogram of the mutual inclination between the orbit of HR 2562 B and its disk. The distribution peaks median value of $3.70^\circ$, indicating that HR 2562 B's orbit is nearly coplanar with the disk.}
    \label{fig:mutual_inc}
\end{figure}

\section{Results}

\subsection{HR 2562 B}

Our orbit analysis for HR 2562 B revealed strong consistency with the findings of \citet{Zhang_2023}. The argument of periastron ($\omega$) and longitude of the ascending node ($\Omega$) derived in our analysis are in close agreement with the values reported in their study. Specifically, we find a $\omega = 37^{+32}_{-24}$ deg, and $\Omega = 120.00^{+0.24}_{-0.41}$ deg, compared to \citet{Zhang_2023} $\omega = 180^{+70}_{-17}$ deg, and $\Omega = 302\pm 1$ deg. However, to align our $\omega$ and $\Omega$ values, it is necessary to subtract $180^\circ$ from the values of \citet{Zhang_2023}, which accounts for the apparent difference due to our orientation of the z-axis. Our median orbital inclination, $i = 87.00^{+0.20}_{-0.80}$ deg, aligns well with their value of $85\pm 1$ deg, with a difference of $2^\circ$. Both the mass and parallax are driven by the prior which is located in Table  \ref{tab:posterior_results}.

The posterior distribution for eccentricity ($e$) peaks at $0.34^{+0.23}_{-0.18}$, indicating a preference for lower eccentricities. This contrasts with the broader range and higher eccentricities, which peak near 1.0 using only \citet{Zhang_2023, Maire_2018} observations. This result suggests that the orbit of HR 2562 B is less eccentric, with our value being more consistent with a circular orbit. Our eccentricity is significantly lower than that reported in \citet{Zhang_2023}, where the median value is $e=0.80^{+0.20}_{-0.40}$ without enforcing priors due to the disk.

One of the key results for HR 2562 B is the nearly coplanar alignment between the orbit of the companion and the debris disk. Using Equation (8) from \citet{Bean_2009} to calculate the mutual inclination, we find that the orbit is rotated by  $4\pm 1$ deg  from the plane of the disk.  The resulting histogram of the mutual inclination is shown in Figure \ref{fig:mutual_inc}. For this calculation, we used a position angle (PA) of $123\pm 1~^{\circ} $ for the disk midplane \citep[derived by adding $90^\circ$ to the PA of the axis of symmetry reported in ][]{Zhang_2023} and an orbital inclination of $85\pm 5$ deg, both values taken from Table 4 of \citet{Zhang_2023}.  

\subsection{YSES 1 b}

\begin{figure}[h]
    \centering
    \includegraphics[width=\linewidth]{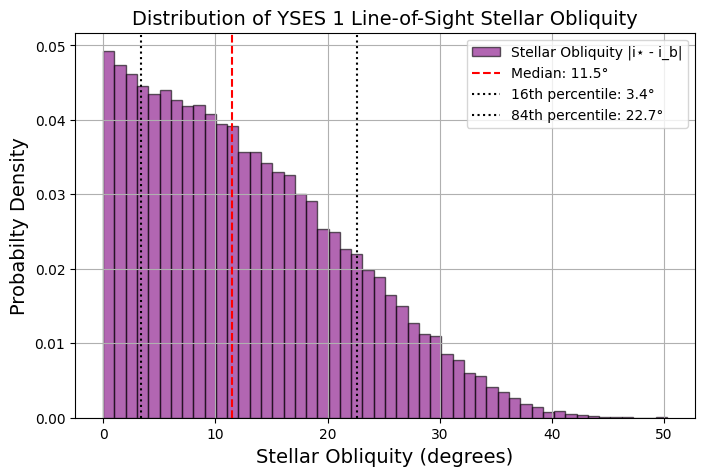}
    \caption{Probability density distribution of the Line-of-Sight stellar obliquity for YSES 1. The median obliquity is 11.5° (red dashed line), with the 16th and 84th percentiles at 3.4° and 22.7° (black dotted lines), respectively.
}
    \label{fig:yses_3}
\end{figure}

The orbital analysis for YSES 1 b reveals an orbit fit driven by the inclusion of planetary radial velocity (RV) data from VLT/CRIRES+ and astrometric data from GRAVITY. The RV measurements, combined with GRAVITY, played a role in constraining the orbit's eccentricity. The posterior distribution shows a  preference for moderate eccentricities with the median of the eccentricity at $0.4\pm 0.2$. 
\begin{figure*}[t]
    \centering
    \includegraphics[width=0.7\linewidth]{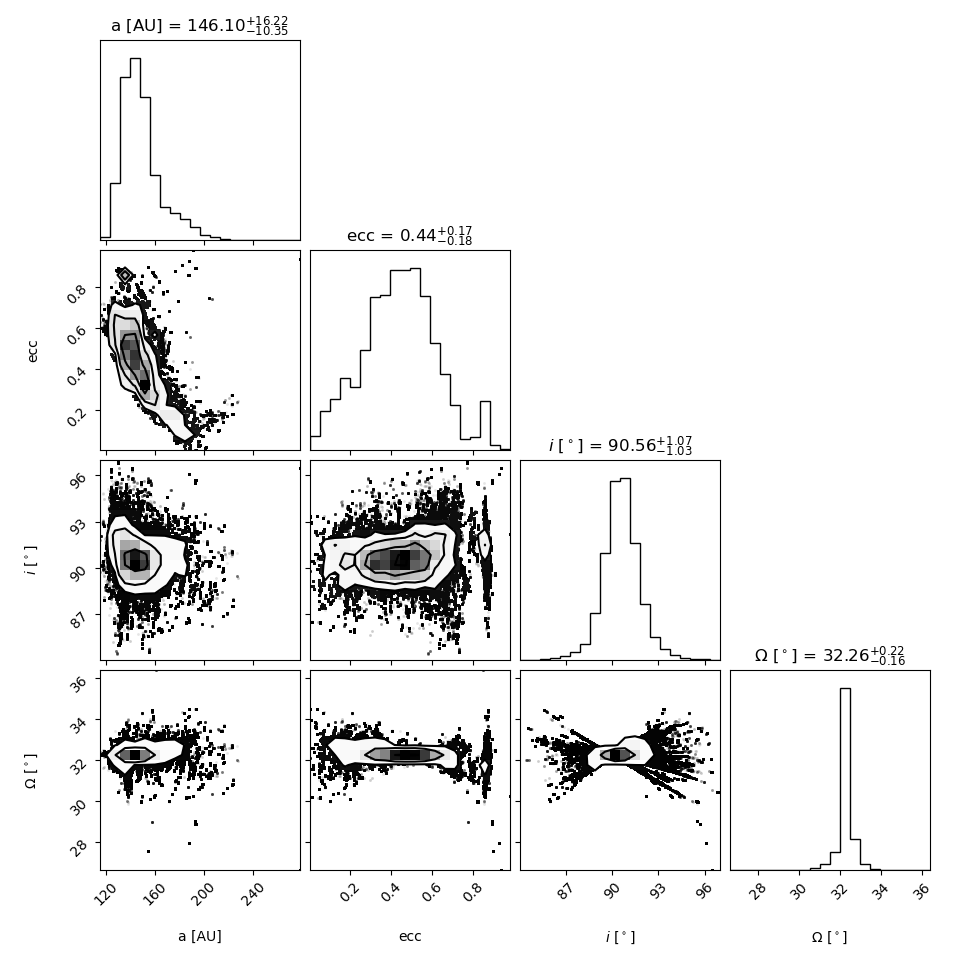}
    \caption{Corner plot for the orbital fit of YSES 1 b. The object demonstrates a preference for eccentric orbits less than 0.5. The semi-major axis distribution shows a prominent peak around 146 AU. }
    \label{fig:yses_small}
\end{figure*}
We determined the line-of-sight stellar obliquity of YSES 1 following the prescription outlined in the appendix of \citet{Bowler_2023}, specifically Equation A12. Our analysis, which incorporates the stellar rotation period from \citet{Bowler_2023}, the stellar radius from \citet{Stassun_2019}, and the projected rotational velocity ($v\sin{i}$) of the star from \citet{zhang2024}, yielded a median line-of-sight stellar obliquity of $12^{+11}_{-8}$° (red dashed line in Figure \ref{fig:yses_3}), with the 16th and 84th percentiles at 3.4° and 22.7°. We also calculate a stellar spin axis line-of-sight inclination median of $78^{+8}_{-11}$°.

\begin{deluxetable*}{lcccc}
    \tablewidth{\textwidth}
    \tablecaption{Orbital Parameter Priors and Posterior Results for YSES 1 b and HR 2562 B
    \label{tab:posterior_results}}
    \tablehead{
        \colhead{\textbf{Parameter}} & 
        \colhead{\textbf{Symbol}} & 
        \colhead{\textbf{Prior}} & 
        \colhead{\textbf{YSES 1 b Posterior}} & 
        \colhead{\textbf{HR 2562 B Posterior}} \\
        \colhead{} & 
        \colhead{} & 
        \colhead{} & 
        \colhead{} & 
        \colhead{}
    }
    \startdata
    Semi-major axis         & $a$                 & Log-uniform prior         & $146^{+16}_{-10}$ AU         & $22.2^{+3.8}_{-2.9}$ AU \\
    Eccentricity            & $e$                 & Uniform [0, 1]            & $0.44^{+0.17}_{-0.18}$        & $0.34^{+0.23}_{-0.18}$ \\
    Inclination             & $i$                 & Sine prior [0°, 180°]     & $90.6^{+1.1}_{-1.0}$ deg     & $86.7^{+0.5}_{-0.7}$ deg \\
    Argument of periastron  & $\omega$            & Uniform [0°, 360°]        & $315^{+8}_{-11}$ deg         & $37^{+32}_{-24}$ deg \\
    Longitude of ascending node & $\Omega$        & Uniform [0°, 360°]        & $32.3^{+0.2}_{-0.2}$ deg     & $119.5^{+0.3}_{-0.7}$ deg \\
    Total System Mass       & $M$                 & Normal prior              & $1.02^{+0.02}_{-0.02}\,M_\odot$ & $1.25^{+0.12}_{-0.09}\,M_\odot$ \\
    Parallax                & $\pi$               & Normal prior              & $10.61^{+0.01}_{-0.01}$ mas  & $29.47^{+0.02}_{-0.02}$ mas\\
    Companion Mass  & $M_{\text{Jup}}$ & Log-uniform prior & — & $\leq 22\,M_{\text{Jup}}$
    \enddata
\end{deluxetable*}

\section{Discussion}


The eccentricty for YSES 1 b suggests a moderately eccentric orbit that does not favor a history of violent scattering or strong dynamical perturbations. The overall distribution is not peaked near $e = 0$, as would be expected from \textit{in situ} formation in a quiescent protoplanetary disk \citep[e.g.,][]{Bowler_2020}. Instead, the moderate eccentricity may point to a more complex dynamical history, potentially involving scattering early in the systems life. 

Using the stability analysis of YSES 1 b \& c from Appendix E of \citet{Bohn_2020}, we find that only $4.20\%$ of our orbits of YSES 1 b are dynamically stable on gigayear timescales with eccentricities less than 0.12. Although \citet{Bohn_2020} assumed planets at 160 and 320 AU, our orbital fits yield semi-major axes that differ from these assumptions, indicating that the previous dynamical modeling may require revision for improved accuracy. Ultimately, a definitive assessment of the system's long-term evolution requires constraints on both the semi-major axis and eccentricity of YSES 1 c, as its orbital parameters remain unknown and could significantly impact the system's overall stability.

The YSES 1 system exhibits a line-of-sight stellar obliquity consistent with 0°. We find no evidence of significant misalignment between YSES 1 b's orbit and the star's spin axis. This alignment could be consistent with a quiescent formation scenario within a protoplanetary disk. If the moderate eccentricity we observe for YSES 1 b were excited through dynamical interactions with the outer companion YSES 1 c, then the low stellar obliquity would suggest that YSES 1 c is also likely to be in a coplanar orbit. Constraining the mutual inclination between YSES 1 b and c would therefore be an important goal for future observations, as it would offer insight into the dynamical history of the system. A similar obliquity analysis was not performed for HR 2562 B, as no measurement of the stellar rotational period is currently available.

While the orbit of YSES 1 b is relatively constrained, with a moderate eccentricity of $0.4\pm 0.2$, the available astrometric data for YSES 1 c are insufficient to meaningfully constrain its orbit, resulting in posterior distributions that remain prior-dominated. This uncertainty allows for several possible scenarios regarding the system's dynamical history. Determining YSES 1 c's orbital parameters, especially its eccentricity and inclination relative to YSES 1 b, is therefore crucial. A high eccentricity for YSES 1 c and/or a significant mutual inclination would suggest past dynamical interactions, such as planet-planet scattering, that may have excited the orbital eccentricities of one or both companions. Conversely, low values for YSES 1 c's eccentricity would support a more quiescent evolution.

The $4\pm 1$ deg offset observed between the orientation of HR 2562 B's orbit and the disk provides important clues about the companion's formation and evolutionary history. This small mutual inclination, while non-zero, indicates that the orbit of HR 2562 B is nearly coplanar with the debris disk. Given the moderate eccentricity of $0.4\pm 0.2$, and the co-planarity of the orbit with the debris disk, it is possible that any dynamical excitations that HR 2562 B might have experienced could have occurred within the plane of the disk. However, the offset also raises the possibility of mild dynamical interactions, such as planet-planet scattering or perturbations from an unseen companion, which could have slightly tilted the orbit relative to the disk. 

Additionally, we calculate an apastron of $29^{+4}_{-2}$ AU, with a maximum of 58.5 AU, which is consistent with the limits of the debris disk inner edge of 63 AU from \citet{zhang2024}.  The companion mass we derive has an upper limit of $22\,M_{\text{Jup}}$ and falls within the lower limit of the estimated mass of $30 \pm 15\, M_{\text{Jup}}$ \citep{KONO2016}. This constraint puts HR 2562 B at the lower end of the age range of 300-900 Myr.

\begin{acknowledgements}

In this work we utilized the Python \citep{python} and Julia \citep{julia} programming languages. Notably we used Numpy \citep{numpy_cite},  Orbitize \citep{orbitize}, Pandas \citep{pandas}, Astropy \citep{astropy_2013,astropy_2018}, PyMC \citep{pymc}, and Corner \citep{corner}.

J.R. expresses gratitude to the Illinois/NASA Space Grant Consortium for their support, which, while not directly resulting in a publication, provided valuable assistance in advancing my academic career.
J.R. thanks the LSST-DA Data Science Fellowship Program, which is funded by LSST-DA, the Brinson Foundation, the WoodNext Foundation, and the Research Corporation for Science Advancement Foundation; his participation in the program has benefited this work. J.J.W. and J.R. are supported by NASA XRP Grant 80NSSC23K0280. A.Z. acknowledges support from ANID -- Millennium Science Initiative Program -- Center Code NCN2024\_001.

\end{acknowledgements}

\clearpage

\begin{figure}[h]
    \centering
    \includegraphics[width=\linewidth]{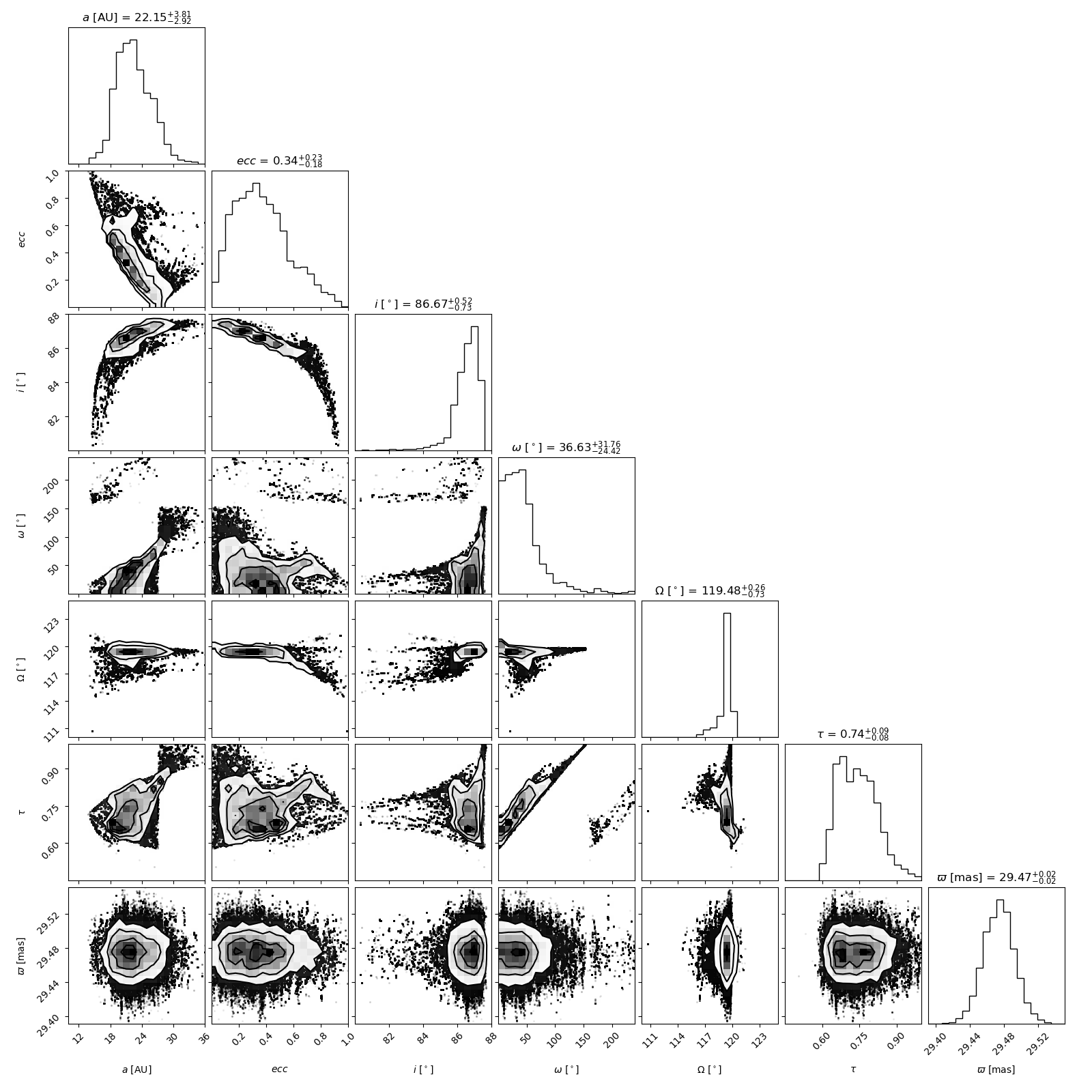}
    \caption{Full Corner plot of HR 2562 B. The estimated posteriors are on the diagonals. The cross-sections represent the covariance with the 1, 2, and 3-$\sigma$ errors.
}
    \label{fig:HR2562_large}
\end{figure}

\begin{figure}[h]
    \centering
    \includegraphics[width=\linewidth]{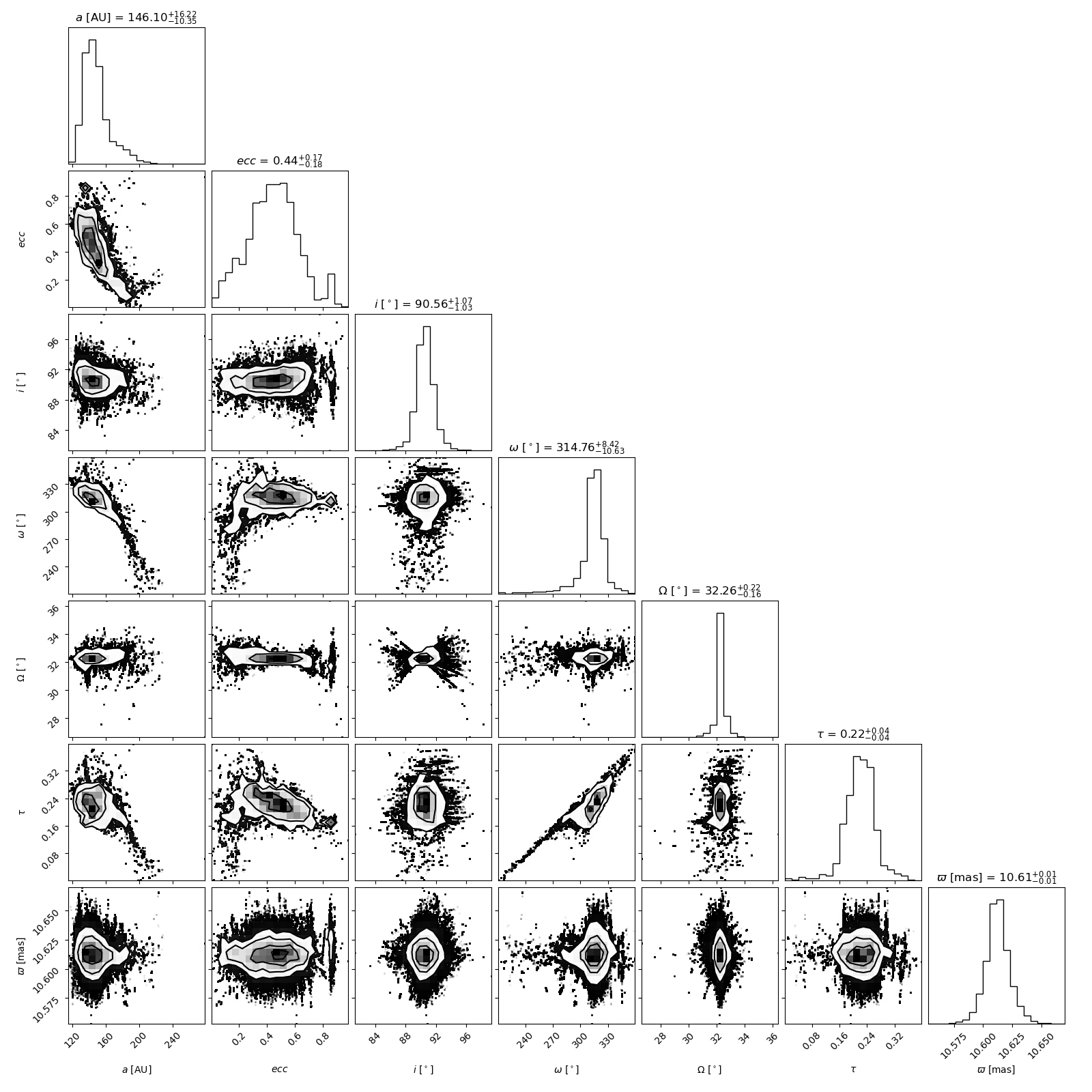}
    \caption{Full corner plot for the orbital fit of YSES 1 b.}
    \label{fig:yses_large}
\end{figure}

\FloatBarrier
\appendix
In figures \ref{fig:HR2562_large} and \ref{fig:yses_large} show the full posterior distributions of both HR 2562 B and YSES 1 b. 

\clearpage

\bibliography{orbits}
\bibliographystyle{aasjournal}
\bibliographystyle{yahapj.bst}


\end{document}